\documentstyle[twoside,fleqn,espcrc2,epsf]{article}
\title{
Calculation of $K\rightarrow \pi$ matrix elements in quenched domain-wall QCD 
\thanks{Talk presented by J.~Noaki} }

\author{CP-PACS Collaboration :
  A.~Ali~Khan\rlap,\address{Center for Computational Physics,
    University of Tsukuba, Tsukuba, Ibaraki 305-8577, Japan}
  S.~Aoki\rlap,\address{Institute of Physics,
    University of Tsukuba, Tsukuba, Ibaraki 305-8571, Japan}
  Y.~Aoki\rlap,$^{\rm a,b}$\thanks{address after 1 May, 2000:
        RIKEN BNL Research Center, Brookhaven National
        Laboratory, Upton, NY 11973, USA}
  R.~Burkhalter\rlap,$^{\rm a,b}$
  S.~Ejiri\rlap,$^{\rm a}$
  M.~Fukugita\rlap,\address{Institute for Cosmic Ray Research,
    University of Tokyo, Kashiwa 277-8582, Japan}
  S.~Hashimoto\rlap,\address{High Energy Accelerator Research Organization
    (KEK), Tsukuba, Ibaraki 305-0801, Japan}
  N.~Ishizuka\rlap,$^{\rm a,b}$
  Y.~Iwasaki\rlap,$^{\rm a,b}$
  T.~Izubuchi\rlap,\address{Institute of Theoretical Physics, Kanazawa
    University, Ishikawa 920-1192, Japan}
  K.~Kanaya\rlap,$^{\rm b}$
  T.~Kaneko\rlap,$^{\rm d}$
  Y.~Kuramashi\rlap,$^{\rm d}$
  K.-I.~Nagai\rlap,$^{\rm a}$
  J.~Noaki\rlap,$^{\rm a}$
  M.~Okawa\rlap,$^{\rm d}$
  H.P.~Shanahan\rlap,$^{\rm a}$\thanks{address after 15 Sept., 2000:
        Department of Biochemistry and Molecular
        Biology, University College London, London, England, UK}
  Y.~Taniguchi\rlap,$^{\rm b}$
  A.~Ukawa$^{\rm a,b}$ and
  T.~Yoshi\'e$^{\rm a,b}$
  }

\begin{document}

\begin{abstract}
We explore the possibility for an evaluation of 
non-leptonic $\Delta S=1$ $K$ decay amplitudes through the calculation
of $K\to\pi$ matrix elements using domain-wall QCD.  
The relation between the physical $K\to\pi\pi$ matrix 
elements and $K\to\pi$ matrix elements deduced from chiral perturbation 
theory is recapitulated. 
Quenched numerical simulations are performed 
on an $16^3\times 32\times 16$ lattice 
at a lattice  spacing $a^{-1}\approx 2$GeV 
for both the standard plaquette gauge action and a 
renormalization-group improved gauge action,
and reasonable signals for $K\to\pi$ matrix elements are obtained.
Preliminary results are reported on 
the $K\to\pi\pi$ matrix elements,
and results from two actions are compared. 
\end{abstract}
\maketitle

\section{INTRODUCTION}

Despite extensive efforts over the years, lattice QCD calculation of 
the matrix elements relevant for understanding the $K\to\pi\pi$ decays, 
including the $\Delta I=1/2$ rule and the direct $CP$ violation 
parameter $\varepsilon^\prime/\varepsilon$,  
have achieved only limited success to the present\cite{lellouch}.
One reason behind the slow progress is lack of full chiral symmetry in the 
Kogut-Susskind\cite{PK} and Wilson\cite{JLQCD} formulation 
for lattice fermions employed in 
the past attempts.  The domain wall fermion formalism \cite{Kaplan,FS} 
offers a possibility of resolving this problem.  
In this article we report on our study of the $K\to\pi\pi$ 
amplitudes in quenched QCD using this formalism\cite{columbia}. 

The framework of our study is the reduction formulae\cite{Bernard85} 
derived from chiral perturbation theory($\chi$PT), 
which relate $K\to\pi$ amplitudes to the physical $K\to\pi\pi$ amplitudes. 
We briefly review these formulae to expose some 
limitations\cite{who,ishizuka}. 

Chiral property of domain wall QCD has been examined in detail recently
\cite{CPPACS_orig,RBCcollab}.  We have shown in particular
\cite{CPPACS_orig} that the residual chiral symmetry breaking due to 
finite fifth-dimensional size is significantly reduced for a 
renormalization-group improved gluon action as compared to the standard 
plaquette action. We therefore calculate the $K\to\pi$ matrix elements 
for the two gluon actions in parallel.

\section{$\chi $PT REDUCTION FORMULAE }
The effective Hamiltonian $H_W$ for $\Delta S=1$ K meson decays 
is written as 
\begin{eqnarray}
 H_W^{\rm (eff)}  =\frac{G_F}{\sqrt{2}}V_{ud}V_{us}^*\sum_{i=1}^{10}
  W(\mu)_iQ_i(\mu),
\end{eqnarray}
where $W(\mu)_i$ are Wilson coefficients. The basis four-quark operators 
$Q_i$ are decomposed as $ Q^{(0)}_i+Q^{(2)}_i$ corresponding 
to iso-spin in the final state $I=0$ and $2$.

Under $SU(3)_L\otimes SU(3)_R$ chiral transformation,
the basis operators $Q_i$ transform as $(27,1)\oplus(8,1)$ for $i=1,
\dots,4,9,10$, $(8, 1)$ for $i=5,6$ and $(8,8)$ for $i=7,8$.
To leading order in $\chi$PT the operators that transform 
according to these flavor representations are given 
by\cite{Bernard85,who,ishizuka}
\begin{eqnarray}
(8,1):\ &{\bf A}&=(\partial_\mu\Sigma^\dagger\partial_\mu\Sigma)_{23},\\
        &{\bf B}&=(\Sigma^\dagger M+M^\dagger\Sigma)_{23}\\
(27,1):\ &{\bf C}&=3(\Sigma^\dagger\partial_\mu\Sigma)_{23}     
                    (\Sigma^\dagger\partial_\mu\Sigma)_{11} \nonumber \\
                & &\ \ +2(\Sigma^\dagger\partial_\mu\Sigma)_{13}       
                    (\Sigma^\dagger\partial_\mu\Sigma)_{21}\\
(8,8):\ &{\bf D}&=\Sigma^\dagger_{21}\Sigma_{13},
\end{eqnarray}          
where we consider $(8,8)$ operator only at $O(p^0)$ order due to the
reason explained later.

Using unknown parameters, $a,b,c,d\dots$, one may write the $Q_i$'s  
in terms of the $\chi$PT operators according to 
\begin{eqnarray}
Q_i\!\!&=&\!\!a_i{\bf A}+b_i{\bf B}+c_i{\bf C}\ \  (i=1,\dots,6,9,10)
\label{relation-1}\\
Q_i\!\!&=&\!\!d_i{\bf D}\ \ (i=7,8)
\label{relation-2} 
\end{eqnarray}
for the case of $I=0\ {\rm and}\ 2$ separately. 

For $Q_i$ in (\ref{relation-1}), 
$K^+\to\pi^+$ matrix elements are proportional to
$a_i+b_i-2c_i$, whereas $K^0\to\pi^+\pi^-$ matrix elements to $a_i-2c_i$.
To eliminate $b_i$ one has to introduce an operator
\begin{eqnarray}
Q_{\rm sub}=(m_s+m_d)\bar{s}d-(m_s-m_d)\bar{s}\gamma_5d = b_{\rm sub}{\bf B}
\end{eqnarray}
and construct $Q_i - \alpha_i Q_{\rm sub}$, where $\alpha_i =b_i/b_{\rm sub}$
can be determined by $K^0\to 0$ matrix elements (See (\ref{alpha_def})).
One then obtains the reduction formula,  (\ref{origred}) below,
relating the $K\to\pi\pi$ and $K\to\pi$ amplitudes originally given by 
Bernard {\it et.al}\cite{Bernard85}.

In (\ref{relation-2}) there would appear too many unknown parameters 
to invalidate the reduction relation,
if all $(8,8)$ operators  up to ${\cal O}(p^2)$ order 
were included as in the case of other representation. 
To exclude contributions from these operators  at $O(p^2)$,
we take the chiral limit 
where the reduced relation is justified.
In this limit, the matrix elements for $K\to\pi\pi$ are obtained 
in proportion to those for $K\to\pi$\cite{who,ishizuka}.

It should be stressed that it is only in the chiral limit that one can 
obtain a complete set of reduction formulae at the lowest order of $\chi$PT.

\section{NUMERICAL SIMULATION}

\begin{table}[h]
\caption{Simulation parameters. Scale $a^{-1}$ is fixed from string 
tension $\sqrt{\sigma}=440$~MeV.} 
\label{params}
\begin{tabular}{|c|c|c|}
\hline 
gauge action & plaquette & RG-improved\\
\hline\hline
$\beta$ & 6.0  &2.6\\
\hline
$a^{-1}$ &2.0 GeV  &1.94 GeV\\
\hline
lattice size &\multicolumn{2}{|c|}{$16^3\times 32\times16$}\\
\hline
DW height $M\!\!\!\!$ &\multicolumn{2}{|c|}{1.8}\\
\hline
$m_f$ & \multicolumn{2}{|c|}{$0.01-0.06$ in steps of 0.01}\\
\hline
\#config. & 111 &110 \\
\hline\hline
$m_{5q}$\cite{CPPACS_orig}
& 2.96(34) MeV\hspace{-1mm} & 0.283(42) MeV\hspace{-4mm}\\
\hline
\end{tabular}
\vspace{-15pt}
\end{table}

Parameters of our numerical simulation are summarized in 
Table.~\ref{params}.  
Calculations are carried out, assuming degenerate bare quark mass 
$m_f=m_u=m_d=m_s$, on a $16^3\times 32\times 16$ lattice at a 
coupling constant corresponding to a lattice spacing of 
$a^{-1}\approx 2$~GeV.  
With these parameters the anomalous quark mass $m_{5q}$ representing 
residual chiral symmetry breaking is similar to u-d quark mass for the 
plaquette action and quite small for the RG-improved action. 
We set the scale from measurement of the string tension assuming 
$\sqrt{\sigma}=440$~MeV\cite{CPPACS_orig}. 

To evaluate quark loops in some types of contractions,
we employ the random noise method with the number of noises taken to be 2.  
This number is decided from a numerical test.
 
We calculate the $K\to\pi$ matrix elements using the wall source and 
dividing by the normalization factor
$\left<\pi^+\left|A_4\right|0\right>\left<0\left|A_4\right|K^+\right>$.
In this convention, $\chi$PT reduction formulae for $i=1,\dots,6,9,10$
take the form, 
\begin{eqnarray}
\left<\pi^+\pi^-\left|Q_i^{(I)}\right|K^0\right>
\!\!\!\!& &=\sqrt{2}f_\pi(m_K^2-m_{\pi}^2) \nonumber\\ 
\times\ & &\!\!\!\!\!\!\!\!\!\!\!\!\!\!
\frac{\left<\pi^+\left|Q_i^{(I)}-\alpha_i
Q_{\rm sub}\right|K^+\right>}{\left<\pi^+\left|A_4\right|0\right>
        \left<0\left|A_4\right|K^+\right>},\label{origred}
\end{eqnarray}
\vspace{-2mm}
and for $i=7,8$,
\begin{eqnarray}
\left<\pi^+\pi^-\left|Q_i^{(I)}\right|K^0\right>
\!\!\!\!& &=\nonumber \\
\!\!\!\! -\sqrt{2}f_\pi  \times& &\!\!\!\!\!\!\!\!\!\!\!
m_M^2\frac{\left<\pi^+\left|Q_i^{(I)}\right|K^+\right>}
{\left<\pi^+\left|A_4\right|0\right>\left<0\left|A_4\right|K^+\right>},
\label{newred}
\end{eqnarray}
where $m_M$ is the meson mass on the lattice, while the experimental 
values are to be substituted in $f_\pi, m_\pi, m_K$.
In (\ref{origred}) the subtraction term $\alpha_iQ_{\rm sub}$ appears 
only for $I=0$. 
The parameter $\alpha_i$ is written as
\begin{eqnarray}
\alpha_i= -\frac{\frac{d}{dm_s}\left<0\left|Q_i^{(0)}\right|K^0\right>
|_{m_s=m_d}}{\left<0\left|\bar{s}\gamma_5d\right|K^0\right>},
\label{alpha_def}
\end{eqnarray}
where the differentiation by $m_s$ is implemented by 
a double inversion of the Dirac operator.

\section{RESULTS}

\begin{figure}[t]
  \vspace{-20pt}
  \begin{center}
   \leavevmode
\hspace{-3mm}
  \epsfxsize=3.98cm \epsfbox{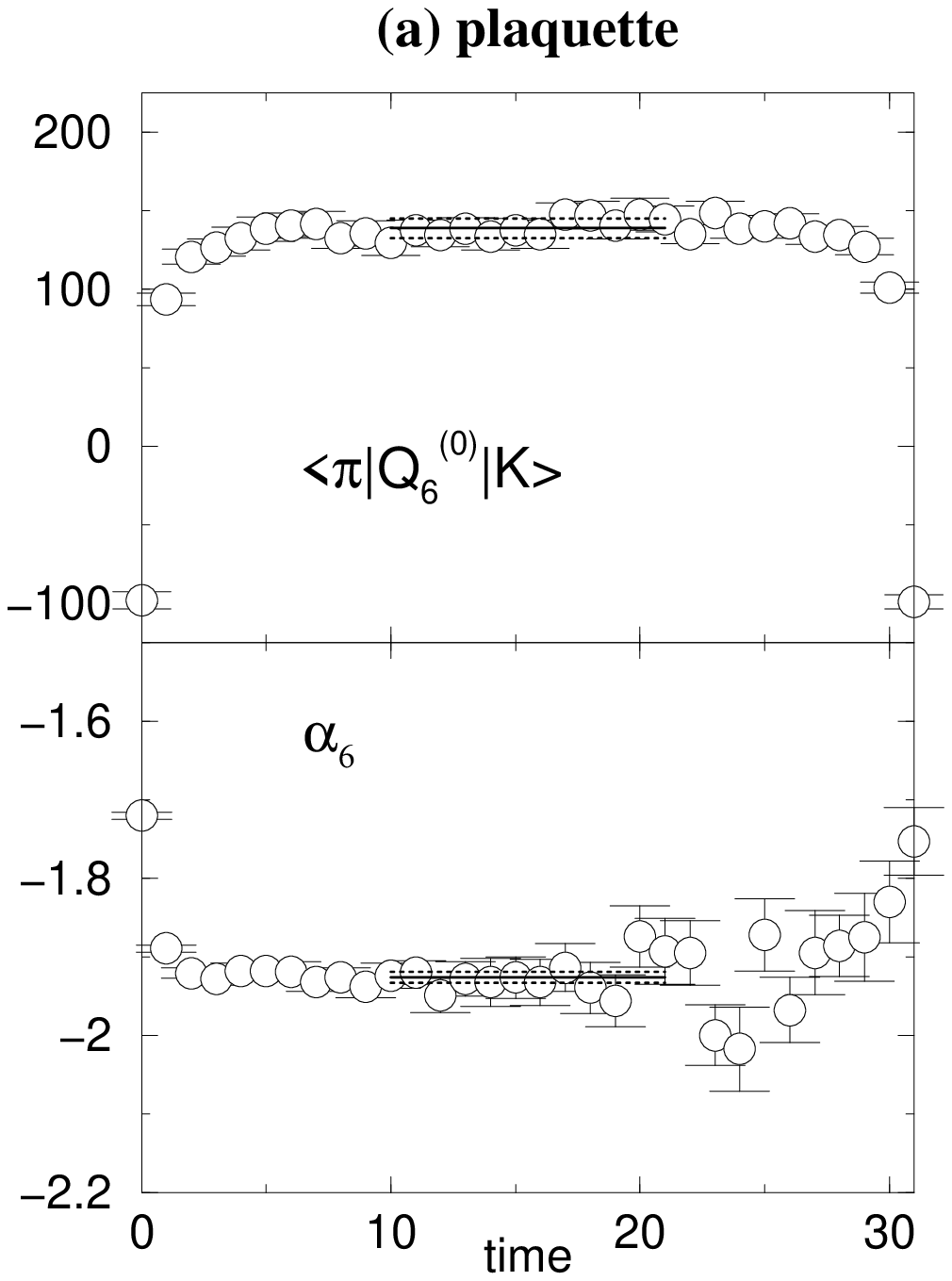}
\hspace{0.3mm}
  \epsfxsize=3.4cm \epsfbox{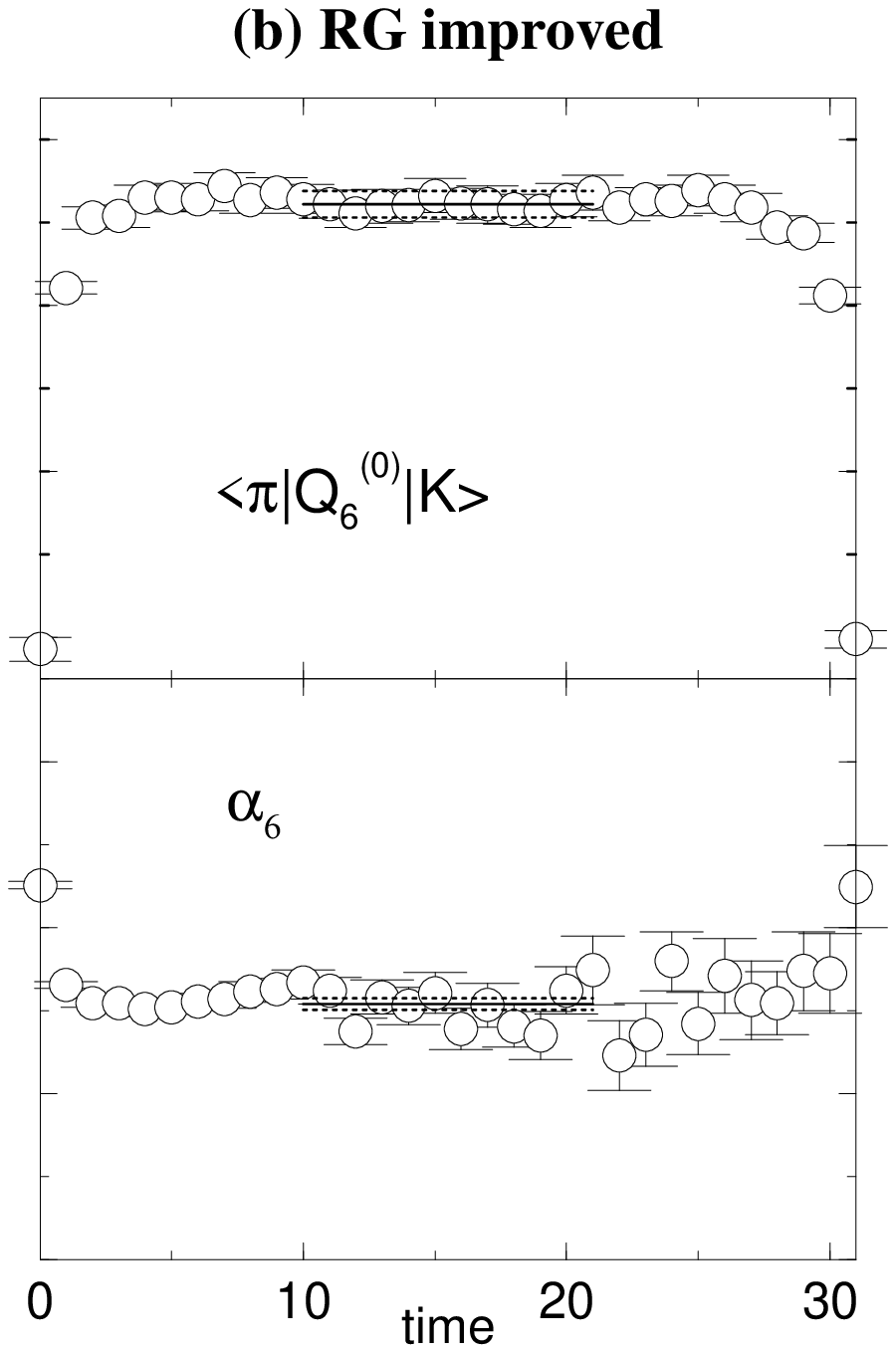}
\hspace{-2.5mm}
  \end{center}
  \vspace{-38pt}
  \caption{Propagators relevant to $ Q_6^{(0)}$ with $m_f=0.03$ 
        for (a) plaquette and (b) RG-improved gauge actions. Fit ranges
        and jack knife errors are represented by horizontal lines. }
  \label{t-dep06}
  \vspace{-10pt}
\end{figure}

\begin{figure}[t]
  \vspace{-20pt}
  \begin{center}
   \leavevmode
\hspace{-5mm}
  \epsfxsize=4.11cm \epsfbox{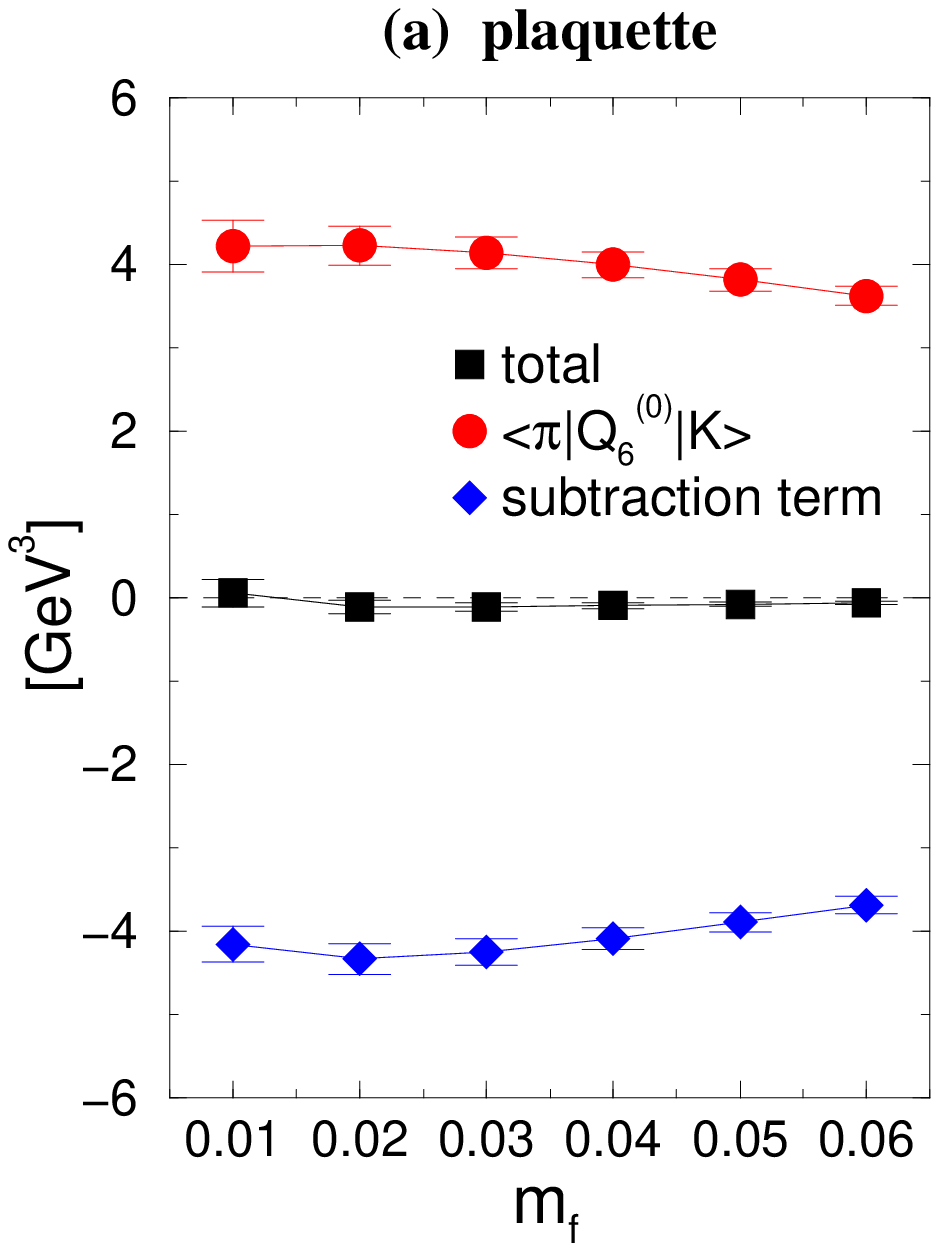}
\hspace{-1mm}
  \epsfxsize=3.59cm \epsfbox{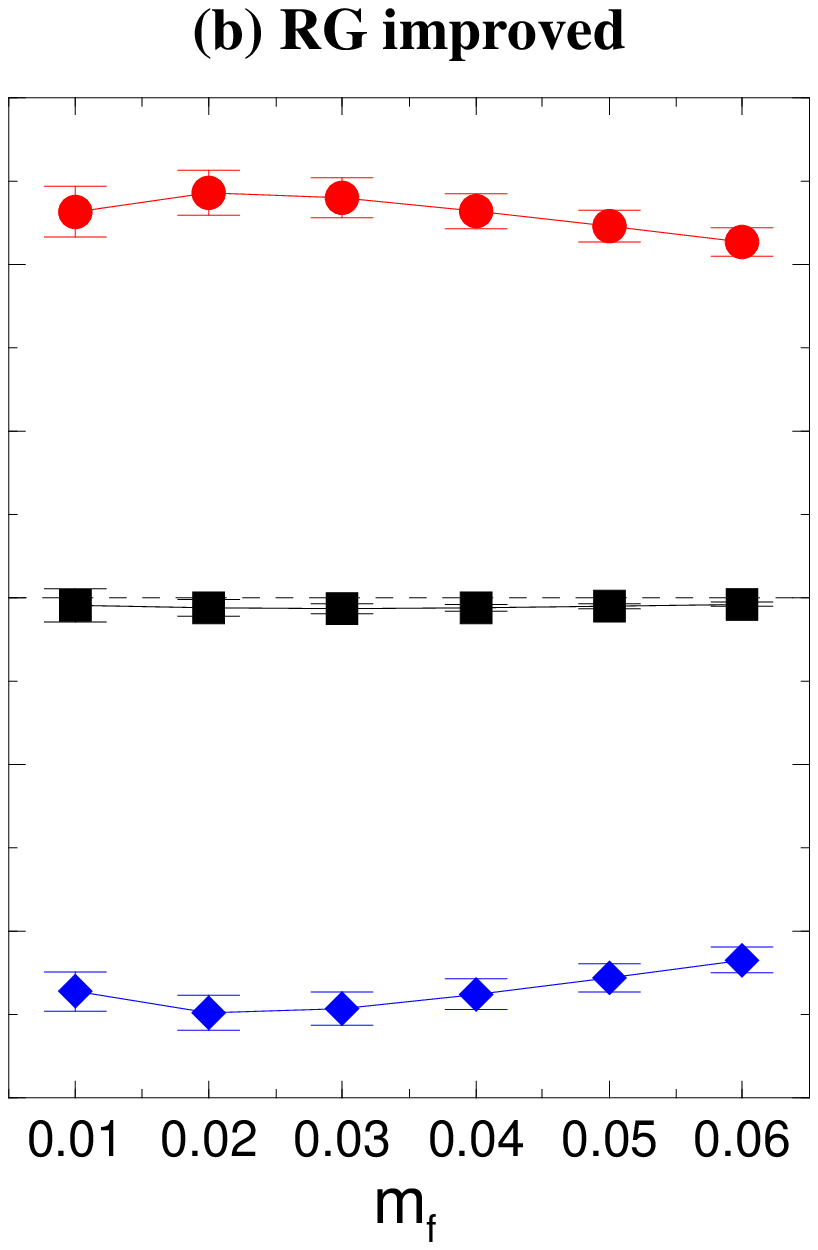}
  \end{center}
  \vspace{-38pt}
  \caption{$Q^{(0)}_6$ calculated by the $\chi$PT reduction formula 
        (\ref{origred}).
        Original matrix element (circles), subtraction term (diamonds)
        and the total value (squares) are separately plotted. Lines are guide
        to eyes.}
  \label{IT06}
  \vspace{-15pt}
\end{figure}

\begin{figure*}
  \vspace{-25pt}
  \begin{center}
  \leavevmode
  \epsfxsize=6.5cm \epsfbox{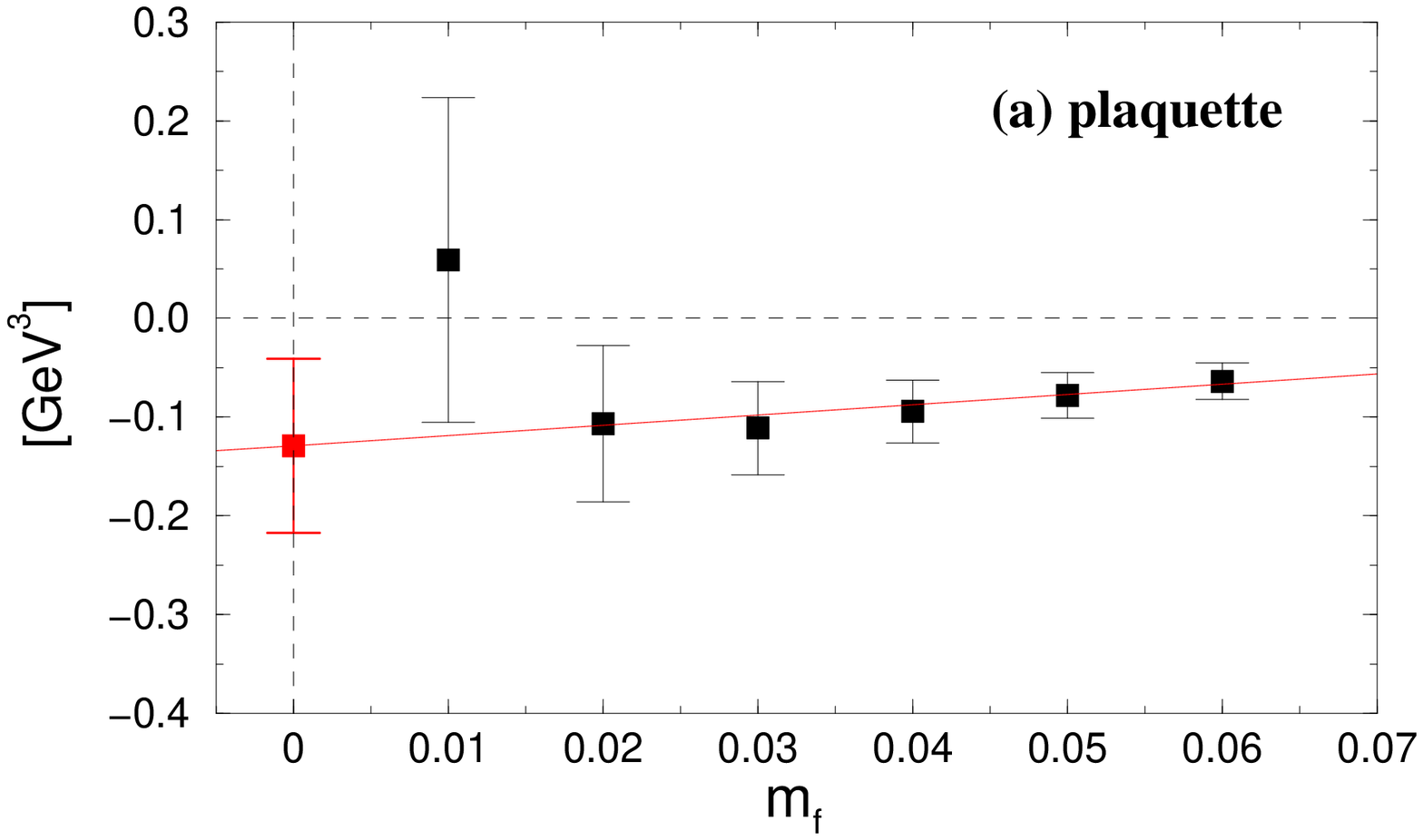}
\hspace{2mm}
  \epsfxsize=6.5cm \epsfbox{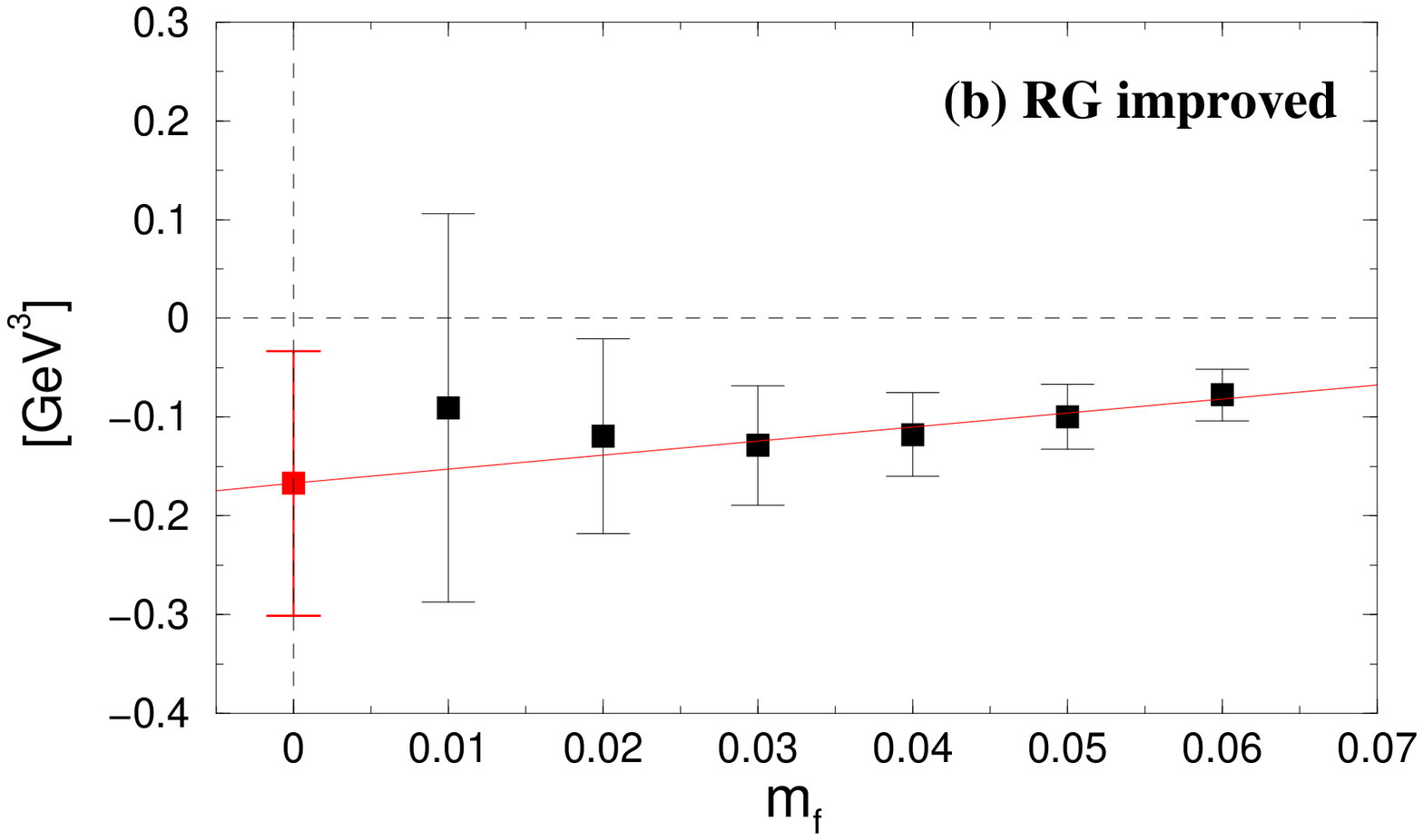}
  \end{center}
  \vspace{-38pt}
  \caption{$K\to\pi\pi$ matrix element of $Q^{(0)}_6$ and linear 
        extrapolation to the limit $m_f\to 0$ for the plaquette 
        and RG-improved gauge action. }
  \label{QL06}
  \vspace{-5pt}
\end{figure*}

\begin{figure*}
  \vspace{-10pt}
  \begin{center}
  \leavevmode
  \epsfxsize=6.4cm \epsfbox{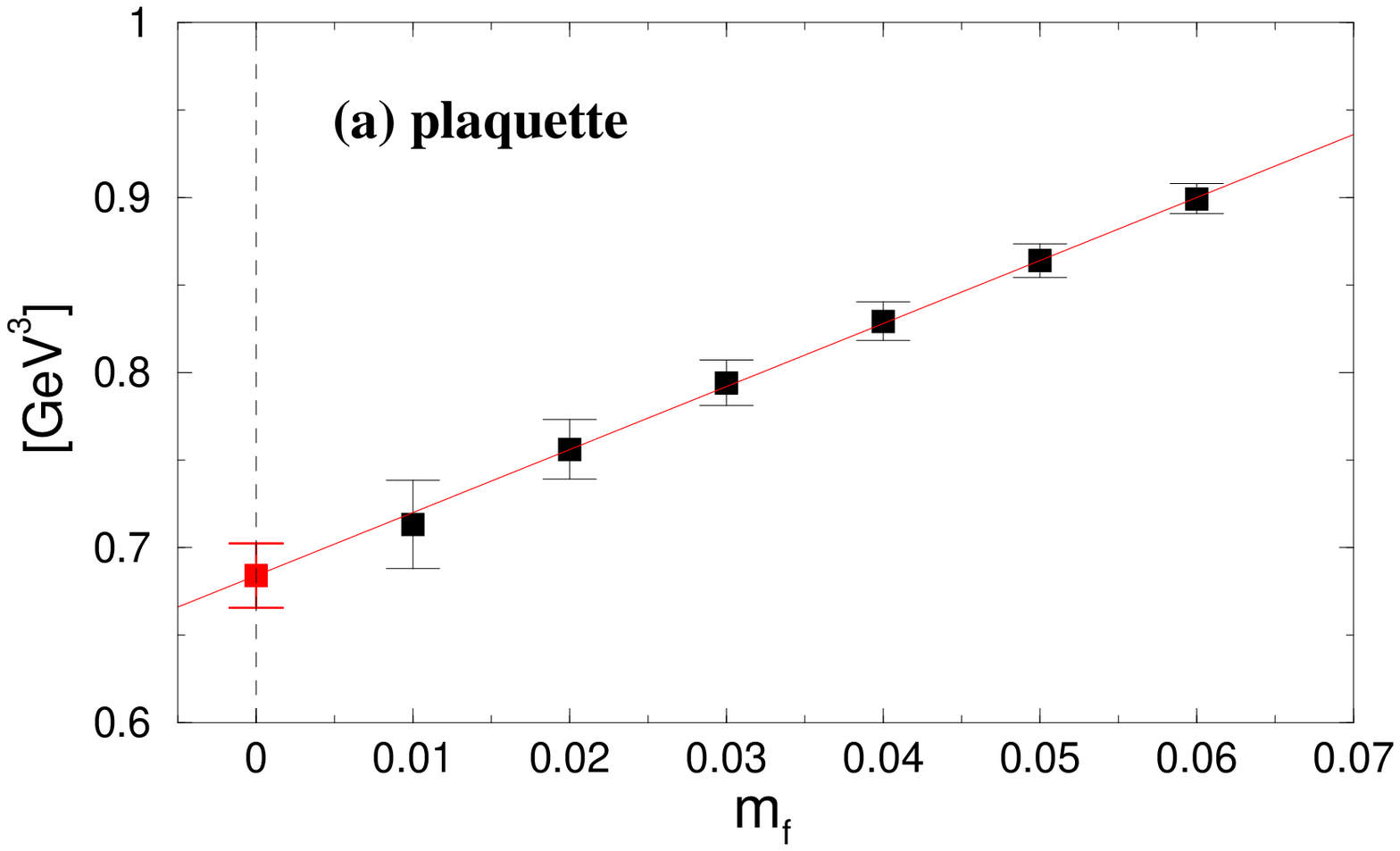}
\hspace{2mm}
  \epsfxsize=6.4cm \epsfbox{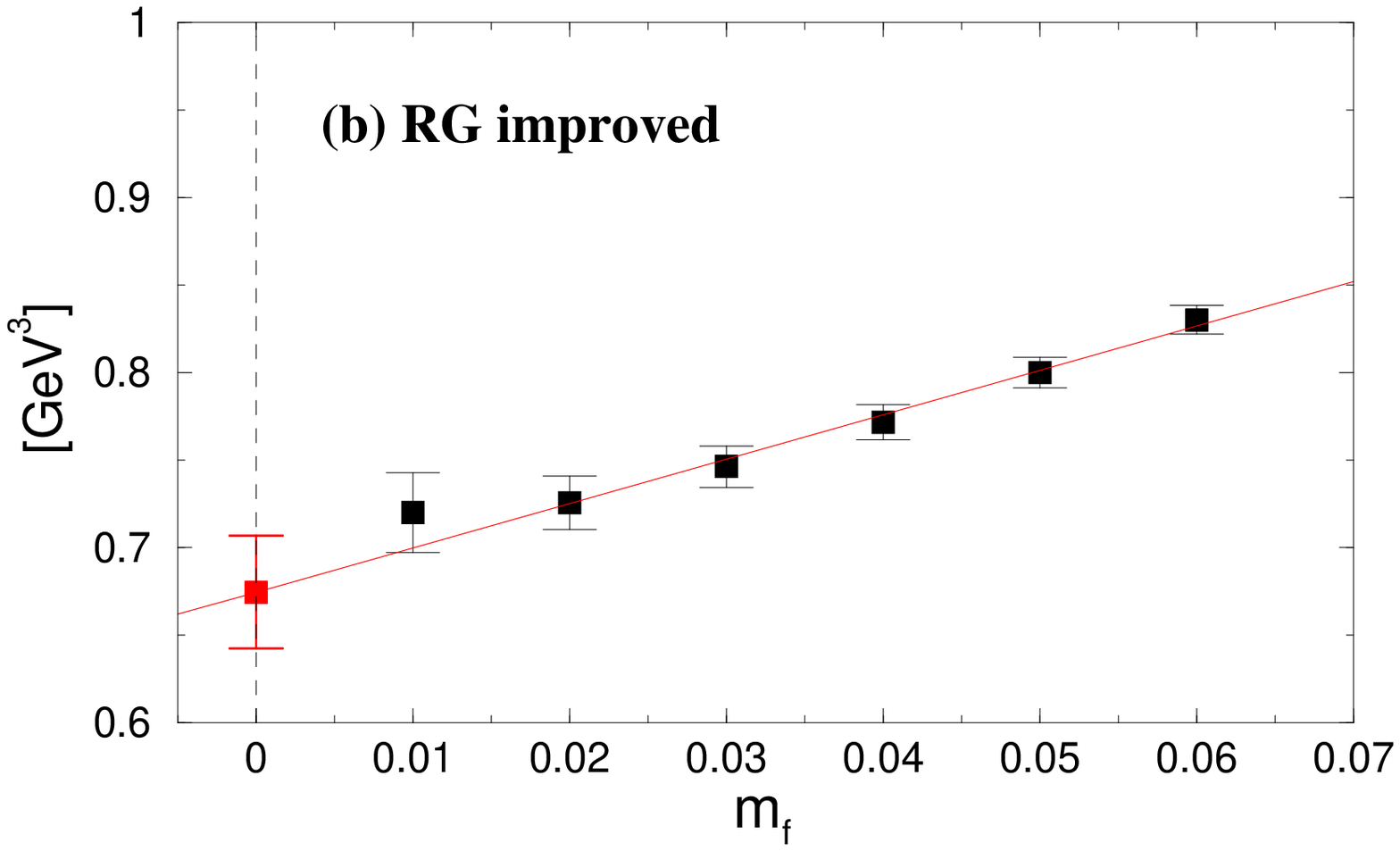}
  \end{center}
  \vspace{-38pt}
  \caption{Counterpart of Fig.~\ref{QL06} for the operator $Q^{(2)}_8$.}
  \label{QL28}
  \vspace{-10pt}
\end{figure*}

For limitation of space we concentrate on the matrix elements of 
$Q^{(0)}_6$ and $Q^{(2)}_8$, which represent the main contributions 
to $\varepsilon'/\varepsilon$.

Examples of propagator ratios for $Q_6^{(0)}$ which appears on the right 
hand-side of (\ref{origred}) are shown in 
Fig.~\ref{t-dep06} for (a) plaquette and (b) RG-improved gauge actions. 
In each figure, the upper panel is for the $K\to\pi$ matrix element 
and the lower for $\alpha_6$ defined in (\ref{alpha_def}). 
Although both propagators contain quark loops, which have been known to
make signals worse, reasonable signals are obtained at the level of 
propagator ratios.  
This also justifies our choice of 2 random noises to
evaluate the quark loops.

A constant fitting, as shown in Fig.~\ref{t-dep06}, 
yields the matrix element, which 
we plot as a function of $m_f$ in Fig.~\ref{IT06}. 
Circles and diamonds show the contribution of the 
$K\to\pi$ matrix element and the subtraction term to the $K\to\pi\pi$ 
matrix element of $Q^{(0)}_6$ as given in the reduction formula 
(\ref{origred}).  
There is a severe cancellation between the 
two contributions, leading to the total value of the $K\to\pi\pi$ matrix 
element (squares) which is more than an order of magnitude smaller than 
the individual contributions. 
Nonetheless, the enlarged plots in Fig.~\ref{QL06}(a) and (b) show 
that the total values have a reasonable signal.  

There is no subtraction term for $Q^{(2)}_8$.  We obtain clear signals 
for this operator using the reduction formula (\ref{newred}) as shown 
in Fig.~\ref{QL28}. 

The remaining procedure 
is to make a linear fitting, as shown in Figs.~\ref{QL06} and \ref{QL28}, 
to take the chiral limit $m_f\to 0$ where both of the reduction formulae 
(\ref{origred}) and (\ref{newred}) are valid to estimate the physical 
matrix element, 
and to convert the values to continuum theory renormalized 
in the $\overline{\rm MS}$ scheme with NDR.  
For the latter, we employ the renormalization factors calculated 
in perturbation theory at one-loop level\cite{renorm} 
with mean field improvement.

Prior to this final step, we are currently increasing the statistics to 
reduce the errors further.  We are also performing a new simulation 
with a larger lattice volume to investigate
finite size effect in the matrix elements.

\section{OUTLOOK}
We have presented our preliminary results for the $K\to\pi\pi$ decay 
amplitudes based on the $\chi$PT reduction formulae and domain wall QCD. 
Matrix elements of reasonable statistical quality are obtained from about 
a hundred gauge configurations in our quenched numerical simulation. 
Values for the matrix elements from plaquette and RG-improved gauge actions 
seems consistent within the errors. These results make us hopeful that 
the present approach yields precise information about the $\Delta I=1/2 $ 
rule and direct CP violation in the standard model with more statistics and 
detailed analysis.

\vspace*{3mm}
This work is supported in part by Grants-in-Aid
of the Ministry of Education (Nos. 
10640246, 10640248, 10740107, 11640250, 11640294, 11740162,
12014202, 12304011, 12640253, 12740133).
AAK is supported by JSPS Research for the Future Program
(No. JSPS-RFTF 97P01102).
SE, TK, KN, JN and HPS are JSPS Research Fellows.

\end{document}